\documentstyle[preprint,aps]{revtex}

\begin{document}
\title{A simple model for DNA denaturation}
\date{\today}
\author{Thomas Garel$^1$, C\'ecile Monthus$^1$ and Henri Orland$^{1,2}$}
\address{
$^1$ CEA, Service de
Physique Th\'eorique, 91191 Gif-sur-Yvette, Cedex, France.\\
$^2$  Institute for Theoretical Physics,
    University of California at Santa Barbara,\\
    Santa Barbara, CA 93106
}
\date{\today}

\maketitle

\bigskip
Pacs: {87.14.Gg, 05.70.Fh, 63.70.+h, 64.10.+h}
\begin{abstract}
 Following Poland and Scheraga, we consider a simplified model for the
denaturation transition of DNA. The two strands are modeled as
interacting polymer chains. The 
attractive interactions, which mimic the pairing between the four bases, are
reduced to a single short range binding term. Furthermore,
base-pair misalignments are forbidden, implying that this binding
term exists only for corresponding (same curvilinear abscissae)
monomers of the two chains. We take into account the excluded volume
repulsion between monomers of the two chains, but neglect
intra-chain repulsion. We find that the excluded volume term
generates an effective repulsive interaction between the chains, which
decays as $1/r^{d-2}$. Due to this long-range repulsion between
the chains, the denaturation transition is first order in any
dimension, in agreement with previous studies.

\end{abstract}
\vskip 10mm
\noindent\mbox{Submitted for publication to: ``Europhys. Lett.''} \hfill 
\mbox{Saclay, SPhT/01-003}\\ \noindent \mbox{ }\\ 
\newpage
The problem of the denaturation of DNA has been of physical interest
for a long time \cite{{Pol_Sch2},{MEF}}, and can be roughly characterized as
the adsorption (or more accurately, desorption) transition of one
strand onto (from) the other. 
Experiments indeed show that the fraction of bound base pairs
exhibits, as a function of temperature, a series of sharp jumps,
linked to the local melting of these heterogeneous paired
structures. This is in contrast with many theories
\cite{{Pol_Sch2},{MEF}}, 
which yield a
continuous denaturation transition below four dimensions.

The full problem is rather complicated
since different binding energies are associated with different
bases contents (such as different AT and GC frequencies).

Moreover, since the denaturation transition decreases the stiffness of
a DNA molecule, the backbone elasticity may play an important
role in the transition \cite{Da_Pe}. Along these lines, the helical
structure of the molecule has been also considered \cite{Co_Mo}

To disentangle polymeric from disorder effects, simplifying
assumptions have to be made. In this work, we follow a time honored
approach where (i) disorder is neglected (homogeneous DNA
approximation) (ii) base-pair mismatch is forbidden (pairing is
possible only between corresponding monomers along the two chains). 
(iii) we consider excluded volume interactions only between the two
chains.

Experimentally, the chains are not very long ($N\approx
1000$) and have a fairly large persistence length ($l_{p}\sim 20$ pairs).
This suggests that for typical distances of the chains
smaller than several persistence lengths, excluded volume effects between 
monomers in the same chain will be negligible compared to excluded volume
effects between monomers of the two chains. 

Closely related work can be found in references
\cite{{Ca_Co_Gr},{Ka_Mu_Pe}} where the existence of a first order
transition in three dimensions has been clearly linked, via numerical
simulations or scaling arguments, to the short
range (excluded volume) repulsion between the chains (by first order,
we mean a phase transition with a finite latent heat and/or some finite
length scales).
\newpage
In our model, the partition function of the two chains of length 
$N$ reads :

\begin{eqnarray}
Z &=&\int {\cal D}\vec{r}_{1}(s)\,{\cal D}\vec{r}_{2}(s)\,\exp \left( -\frac{%
d}{2a^{2}}\int_{0}^{N}ds\,\left( \left( \frac{d\vec{r}_{1}}{ds}\right)
^{2}+\left( \frac{d\vec{r}_{2}}{ds}\right) ^{2}\right) -\beta
\int_{0}^{N}ds\,v(\vec{r}_{1}(s)-\vec{r}_{2}(s))\right)  \nonumber \\
&&\times \exp \left( -g\int_{0}^{N}ds\int_{0}^{N}ds^{\prime } \ \delta
(\vec{r}_{1}(s)-\vec{r}_{2}(s^{\prime }))\right)  \label{partition} 
\end{eqnarray}
where $d$ is the space dimension, $a$ is the Kuhn
length of the monomers, $%
\beta $ is the inverse temperature, $g$ is the excluded volume
parameter and $v(\vec{r}_{1}-\vec{r}_{2})$ is the
short range binding potential for monomers $s$ of chain 1 and 2. In
the following, we will model this interaction by an attractive ($d$ dimensional)
spherical well potential, of radius $r_{0} $ and depth $-V_{0}$. 

An approximate treatment of the excluded volume interaction will be
given first.
Then, we study
the unbinding 
phase transition in the framework of a quantum analogy \cite{DeG}. We
finally compare our results with references \cite{{Ca_Co_Gr},{Ka_Mu_Pe}}.

It is clear from equation (\ref{partition}), that one should focus on 
the relative coordinate $(\vec {r}_1(s)-\vec {r}_2(s))$ of the two
chain system to study a possible adsorption transition. Introducing
the new coordinates
\begin{eqnarray}
\vec{R}(s) &=&\frac{\vec{r}_{1}(s)+\vec{r}_{2}(s)}{2} \\
\vec{r}(s) &=&\vec{r}_{1}(s)-\vec{r}_{2}(s)
\end{eqnarray}

and performing the change of variable in (\ref{partition}), we have

\begin{equation}
Z=\int {\cal D}\vec{r}(s)\,\exp \left( -\frac{d}{4a^{2}}\int_{0}^{N}ds\,%
\left( \frac{d\vec{r}}{ds}\right) ^{2}-\beta \int_{0}^{N}ds\,v(\vec{r}%
(s))-W\left( \left\{ \vec{r}(s)\right\} \right) \right)  \label{relative}
\end{equation}
where

\begin{equation}
e^{-W\left( \left\{ \vec{r}(s)\right\} \right) }=\int {\cal D}\vec{R}%
(s)\,\exp \left( -\frac{d}{a^{2}}\int_{0}^{N}ds\,\left( \frac{d\vec{R}}{ds}%
\right) ^{2}-g\int_{0}^{N}ds\int_{0}^{N}ds^{\prime }\delta \left( \vec{R}(s)-%
\vec{R}(s^{\prime })+{{\vec{r}(s)+\vec{r}(s^{\prime })} \over 2}\right)
\right)
\end{equation}

In the denatured phase, the two chains are far apart and interact
very weakly; it is therefore justified to treat this interaction in a
perturbative manner. Expanding the effective potential $W$ to first
order in $g$, we get

\begin{equation}
e^{-W\left( \left\{ \vec{r}(s)\right\} \right) }\simeq \exp \left(
-g\int_{0}^{N}ds\int_{0}^{N}ds^{\prime }\left\langle \delta \left( \vec{R}%
(s)-\vec{R}(s^{\prime })+\frac{1}{2}(\vec{r}(s)+\vec{r}(s^{\prime }))\right)
\right\rangle _{0}\right)
\end{equation}
where the brackets stand for

\begin{equation}
\left\langle A\right\rangle _{0}=\frac{1}{V}\int {\cal D}\vec{R}%
(s)\,\,\,A\,\exp \left( -\frac{d}{a^{2}}\int_{0}^{N}ds\,\left( \frac{d\vec{R}%
}{ds}\right) ^{2}\right)
\end{equation}
and $V$ is the volume of the system.

We obtain

\begin{equation}
\label{doublev}
W\left( \left\{ \vec{r}(s)\right\} \right) \simeq 
g\int_{0}^{N}ds\int_{0}^{N}ds^{\prime }\left( \frac{ d}{\pi \left|
s-s^{\prime }\right| a^2}\right) ^{d/2}\exp \left( -\frac{d \left( \vec{r}(s)+%
\vec{r}(s^{\prime })\right) ^{2}}{4 a^2 \left| s-s^{\prime }\right| }\right)
\end{equation}
which is valid for dimensions $d>2$. This expression is asymptotically
exact in the denatured phase, where very few monomers of the two
chains come close together.

Setting $S=(s+s^{\prime })/2$ and $\sigma
=s-s^{\prime }$ in equation (\ref{doublev}), we get for large $N$
\bigskip 
\begin{equation}
W\left( \left\{ \vec{r}(s)\right\} \right) \simeq 
g\int_{0}^{N}dS \int_{-\infty }^{+\infty }d\sigma \left( \frac{ d}{\pi \left|
\sigma \right| a^2}\right) ^{d/2}\exp \left( -\frac{d \left( \vec{r}(S+\sigma /2)+%
\vec{r}(S-\sigma /2)\right) ^{2}}{4 a^2 \left| \sigma \right| }\right)
\label{perturb}
\end{equation}

\bigskip

A simple scaling argument shows that this effective interaction must 
vanish at large relative separation.
The range of $|\sigma|$ which contributes to
the integral is given by $|\sigma|_{\rm typ} \sim 
d (\vec{r}(S+\sigma /2)+ \vec{r}(S-\sigma /2)) ^{2}/ {4 a^2} $ . 
Thus,
for small separations where the interaction is sizeable, only small
$\sigma$ contribute significantly to the integral and we may
expand $\ \vec{%
r}(S+\sigma /2)+\vec{r}(S-\sigma /2)$ in powers of $\sigma $, and obtain

\begin{eqnarray}
W\left( \left\{ \vec{r}(s)\right\} \right) &\simeq &
g\int_{0}^{N}dS\int_{-\infty }^{+\infty }d\sigma \left( \frac{ d}{\pi \left|
\sigma \right| a^2}\right) ^{d/2}\exp \left( -\frac{d \vec{r}^{2}(S)}{
a^2 \left|
\sigma \right| }\right) \nonumber \\
&\simeq&\int_{0}^{N}ds\,\frac{\alpha _{d}}{\left| \vec{r}(s)\right|
^{d-2}} \label{effective}
\end{eqnarray}
where
\begin{equation}
\alpha _{d} = 2g \left(\frac{1}{\pi}\right) ^{d/2}\Gamma
(\frac{d}{2}-1){d \over a^2}
\label{alpha}
\end{equation}

We see that the effect of the inter-chain excluded volume
interaction is to generate a ``Coulomb''-like repulsive term in the relative
coordinate. This entropic repulsion is familiar in other interacting
lines problems \cite{MEF,Po_Ta}. Its long range nature is due to the
large lateral excursions of the chains.

 An alternative derivation of
this term can be obtained from equation (\ref{partition}). The use of
the identity  
\begin{equation}
\label{exclu1}
\delta(\vec{r}_{1}(s)-\vec{r}_{2}(s^{\prime }))=\int {d^{d}\vec {k}
\over (2\pi)^{d}} \ e^{i\vec {k} \cdot (\vec{r}_{1}(s)-\vec{r}_{2}(s^{\prime }))}
\end{equation}
allows us to rewrite the excluded volume term of equation
(\ref{partition}) as 
\begin{equation}
\label{exclu2}
{g \over 2}\int_{0}^{N}ds\int_{0}^{N}ds^{\prime }\int {d^{d}\vec {k}
\over (2\pi)^{d}}(\ e^{i\vec {k} \cdot
((\vec{r}_{1}(s)-\vec{r}_{1}(s^{\prime })) 
+(\vec{r}_{1}(s^{\prime})-\vec{r}_{2}(s^{\prime })))} + e^{i\vec {k} \cdot
((\vec{r}_{1}(s)-\vec{r}_{2}(s)
+(\vec{r}_{2}(s))-\vec{r}_{2}(s^{\prime })))})
\end{equation}
where we have explicitly introduced, in a symmetric way, the relative
coordinate of the chains $(\vec {r}_1(s)-\vec {r}_2(s))$.

Each term of equation (\ref{exclu2}) is a product over single chain
and relative coordinate conformations. For each contact of the chains, 
it seems natural to average the single chain term. Defining this
average by $G(\vec k,s,s^{\prime})= 
\langle e^{i\vec {k} \cdot 
(\vec{r}(s)-\vec{r}(s^{\prime }))} \rangle$, the excluded volume term reads
\begin{equation}
\label{exclu3}
g \int_{0}^{N}ds\int_{0}^{N}ds^{\prime }\int {d^{d}\vec {k}
\over (2\pi)^{d}} \ G(\vec k,s,s^{\prime}) \ e^{i\vec k
\cdot (\vec{r}_{1}(s)-\vec{r}_{2}(s))}
\end{equation}
Since we have neglected the intrachain excluded volume, we have
\begin{equation}
G(\vec k,s,s^{\prime})=e^{-{{\vec k}^{2}a^{2} \over 2d} \vert
s-s^{\prime} \vert} 
\end{equation}
Putting everything together, we get back equations
(\ref{effective},\ref{alpha}).

We thus can write the partition function (\ref{relative}) for the relative
coordinate of the two chains as
\begin{equation}
Z=\int {\cal D}\vec{r}(s)\,\exp \left( -\frac{d}{4a^{2}}\int_{0}^{N}ds\,%
\left( \frac{d\vec{r}}{ds}\right) ^{2}-\beta \int_{0}^{N}ds\,v(\vec{r}%
(s))-\int_{0}^{N}ds\,\frac{\alpha _{d}}{\left| \vec{r}(s)\right| ^{d-2}}%
\right)  \label{approx}
\end{equation}

\label{quantik}
For long chains, we may now use a 
well known quantum mechanical analogy \cite{DeG} and study the ground state of
the Hamiltonian
\begin{equation}
\label{Hamilton}
H=-\frac{a^{2}}{d}\vec{\nabla}^{2}+\beta \,v(\vec{r})+\frac{\alpha _{d}}{%
\left| \vec{r}\right| ^{d-2}}
\end{equation}

 This ground state (wavefunction $\Psi_0$, energy $E_0$) is given by
the Schr\"odinger equation 

\begin{equation}
H\,\Psi _{0}=E_{0}\Psi _{0}  \label{schrodinger}
\end{equation}
The free energy per monomer then reads $F=E_{0} \ T$.

We now discuss this equation as a function of space dimensionality.

\begin{itemize}
\item  $2<d<4$

We discuss the typical case $d=3$. The potential is radial, and thus the
ground state will be spherically symmetric.

In radial coordinates, equation (\ref{schrodinger}) becomes

\begin{equation}
\label{radial}
-{a^{2} \over 3} u_{0}^{\prime \prime }+\beta v(r)\,u_{0}+\frac{\alpha
_{3}}{r}\,u_{0}=E_{0}\,u_{0}
\end{equation}
where $u_{0}(r)=r\Psi _{0}(r)$ and $\alpha_3$ is given by (\ref{alpha}).

The unbinding transition \cite{Zi_Li_Kr,Gom} occurs when $T=T_{c}$ and
$E_{0}=0$. As the binding
potential vanishes for $r>r_{0}$, we have in this region

\begin{equation}
-{a^{2}\over 3}u_{0}^{\prime \prime }+\frac{\alpha _{3}}{r}\,u_{0}=0
\end{equation}

The solution of this equation which vanishes at infinity is given by a
Bessel function

\begin{equation}
u_{0}(r) =\sqrt{r \over \lambda}\,K_{1}\left(\sqrt{\frac{r}{\lambda }}\right)
\label{exact} 
\end{equation}
with
\begin{equation}
u_{0}(r)\simeq \left( \frac{r}{\lambda }\right) ^{1/4}\,\exp \left( -\sqrt{\frac{r}{%
\lambda }}\right) \,\,\text{for }r\rightarrow \infty 
\end{equation}
The (finite) correlation length $\lambda$ is given by
\begin{equation}
\lambda =\frac{a^{2}}{12 \alpha_3}= {\pi a^4 \over 72 g}
\end{equation}
implying that the transition is first-order. This is due to the
strongly repulsive barrier generated by the excluded-volume. The
transition temperature $T_{c}\,\ $ is obtained by matching the above
solution to the solution inside the attractive region ($v(r)=-V_0$ for
$r<r_{0}$). The calculations are rather tedious and will not be given
here. It is however interesting to derive an estimate of the latent
heat at the transition. We therefore rewrite equation (\ref{radial})
as 
\begin{equation}
\label{radial2}
-{a^{2} \over 3} u_{0}^{\prime \prime }+\beta_c v(r)\,u_{0}+(\beta -\beta_c)
v(r)\,u_{0}+\frac{\alpha 
_{3}}{r}\,u_{0}=E_{0}\,u_{0}
\end{equation}

Close to $T_{c}$, we do first order perturbation theory and get
\begin{equation}
\label{energie}
E_{0}=\langle u_{0c}(r)\vert (\beta -\beta_c) v(r)\vert
u_{0c}(r)\rangle
\end{equation}
where $ u_{0c}(r)$ denotes the  full solution of equation
(\ref{radial}) at the critical temperature.
Since $F \simeq T_{c} E_{0}$, we finally obtain for the latent heat $L$ 
($L=T_{c} ({\partial F \over \partial T})_{T_{c}}$) 
\begin{equation}
\label{latent}
L=-\langle u_{0c}(r)\vert v(r)\vert
u_{0c}(r)\rangle
\end{equation}

A simpler procedure \cite{DeG} replaces the attractive region by an
attractive $\delta -$potential of strength $\gamma $ at $r=r_0$ leading to

\begin{eqnarray}
F &=&-\gamma \left( \frac{T_{c}-T}{T}\right) \,u_{0}^{2}(r_{0})\text{ for }%
T \to T_{c}^{-} \\
&=&0\text{ for }T>T_{c}
\end{eqnarray}
and a latent heat $L=\gamma u_{0}^{2}(r_{0})$, where $u_0(r_0)$ is
given in equation (\ref{exact}).

\item  $d>4$

The ground state $\Psi _{0}$ satisfies the radial equation

\begin{equation}
-\frac{a^{2}}{d}(\frac{d^{2}}{dr^{2}}+\frac{d-1}{r}\frac{d}{dr})\Psi
_{0}+\left( \beta\,v(r)+\frac{\alpha _{d}}{r^{d-2}}\right) \Psi
_{0}=E_{0}\Psi _{0}
\end{equation}

The unbinding point occurs for $E_{0}=0$\thinspace\ . It is easily seen that
at large distances, the ''Coulomb'' potential is negligible, and thus the
wave-function decays as a power-law

\begin{equation}
\Psi _{0}(r)\sim \frac{1}{r^{d-2}}\,\ \text{for }r\rightarrow \infty
\end{equation}
This wave-function is normalizable for $d>4$, and thus, the transition is
again first-order. This is in agreement with the fact that in $d>4$, the
excluded-volume effect should be irrelevant and thus one should
recover the results of the Gaussian adsorption problem. According to the
Poland-Scheraga theory \cite{Pol_Sch2}, the denaturation transition is
first-order if the exponent $\theta $ describing the 
probability of first return to the origin of the chain is larger than
2. For a Gaussian chain in $d>4$, this is indeed the case since
$\theta ={d \over 2}$.

However, this wave-function is much less localized than below
four dimensions: only moments of $r$ of order strictly smaller than
($d-4$) are finite. 

\item  $d=4$

This case is marginal, since the Laplacian operator and the ''Coulomb''
potential have the same scaling dimension. At large distance, the
wave-function satisfies 
\end{itemize}

\begin{equation}
-\frac{a^{2}}{4}(\frac{d^{2}}{dr^{2}}+\frac{3}{r}\frac{d}{dr})\Psi _{0}+%
\frac{\alpha _{4}}{r^{2}}\Psi _{0}=0
\end{equation}

The wave-function behaves as a power-law

\begin{equation}
\Psi _{0}\sim \frac{1}{r^{\nu }}
\end{equation}
with

\begin{equation}
\nu =1+\sqrt{1+\frac{4 \alpha_4}{a^{2}}}
\end{equation}

The wave-function is normalizable, but again with an infinite correlation
length and only a finite number of moments of $r$ are defined.

We have studied a simple model for the denaturation of DNA, where we
have given an asymptotically exact treatment of the interchain excluded
volume interaction. We find that the first order
character of the denaturation transition stems from the existence of
this long range interaction, in broad agreement with recent work
\cite{Ca_Co_Gr,Ka_Mu_Pe}. The difference in our results might be due
to our neglect of the intrachain excluded volume.
Within their scaling approximation, the
latter authors find that the first order transition is associated with 
a large loop probability distribution, whereas our approximation
yields an exponentially bound state for $d=3$. Future work is needed
to clarify this point. Sequence heterogeneity
\cite{Cu_Hwa}, and chain stiffness \cite{The_Dau_Pey,Bu_La_Li}
raise further questions.

\section*{Acknowledgements}
We would like to thank Y.Kafri and D.Mukamel for many
useful discussions.

\end{document}